\documentclass[universe,article,accept,pdftex,moreauthors]{Definitions/mdpi} 

\usepackage{graphicx}  
\usepackage{amsmath}   
\usepackage{amssymb}   
\usepackage{bm} 
\usepackage{dcolumn}
\usepackage{color}
\usepackage{mathrsfs}
\usepackage{amsfonts}
\usepackage{mathrsfs}
\usepackage{graphicx}
\usepackage{latexsym}
\usepackage{amsmath}
\usepackage{amssymb}
\usepackage{textcomp}
\usepackage{amsbsy}
\usepackage{graphics}
\usepackage{epstopdf}
\usepackage{color}
\usepackage[caption=false]{subfig}
\usepackage{enumitem}
\usepackage{array,multirow} 

\usepackage{float}

\newcommand{\BH}{Bekenstein-Hawking }

\firstpage{1} 
\pubvolume{1}
\issuenum{1}
\articlenumber{0}
\pubyear{2024}
\copyrightyear{2023}
\datereceived{20 November 2023} 
\daterevised{13 December 2023} 
\dateaccepted{19 December 2023} 
\datepublished{ } 
\hreflink{https://doi.org/} 

\Title{Entropic Inflation in Presence of Scalar Field}

\TitleCitation{Entropic Inflation in Presence of Scalar Field}

\Author{Sergei~D.~Odintsov 
 $^{1,2}$, Simone~D'Onofrio $^{2}$ and Tanmoy~Paul 
 $^{3,4,}$*}

\AuthorNames{Sergei~D.~Odintsov, Simone~D'Onofrio and Tanmoy~Paul}

\AuthorCitation{Odintsov, S.D.; D'Onofrio, S.; Paul, T.}

\address{%
$^{1}$ \quad ICREA, Passeig Luis Companys, 23, 08010 Barcelona, Spain; odintsov@ice.csic.es\\
$^{2}$ \quad Institute of Space Sciences (ICE, CSIC) C. Can Magrans s/n, 
 08193 Barcelona, Spain; donofrio@ice.csic.es\\
$^{3}$ \quad Department of Physics, Visva-Bharati University, Santiniketan 731235, 
 India\\
$^{4}$ \quad Labaratory for Theoretical Cosmology, International Centre of Gravity and Cosmos,
Tomsk State University of Control Systems and Radioelectronics (TUSUR), 634050 Tomsk, Russia}

\corres{Correspondence: pul.tnmy9@gmail.com 
}

\abstract{In spirit of the recently proposed four-parameter generalized entropy of apparent horizon, we investigate inflationary cosmology where the matter field inside of the horizon is dominated by a scalar field with a power law potential (i.e., the form of $\phi^n$ where $\phi$ is the scalar field under consideration). Actually without any matter inside of the horizon, the entropic cosmology leads to a de-Sitter spacetime, or equivalently, an eternal inflation with no exit. Thus in order to achieve a viable inflation, we consider a minimally coupled scalar field inside the horizon, and moreover, with the simplest quadratic potential. It is well known that the $\phi^2$ potential in standard scalar field cosmology is ruled out from inflationary perspective as it is not consistent with the recent Planck 2018 data; (here it may be mentioned that in the realm of ``apparent horizon thermodynamics'', the standard scalar field cosmology is analogous to the case where the entropy of the apparent horizon is given by  the Bekenstein--Hawking entropy). However, the story becomes different if the horizon entropy is of generalized entropic form, in which case, the effective energy density coming from the horizon entropy plays a significant role during the evolution of the universe. In particular, it turns out that in the context of generalized entropic cosmology, the $\phi^2$ potential indeed leads to a viable inflation (according to the Planck data) with a graceful exit, and thus the potential can be made back in the scene.}


\begin{document}

\section{Introduction}

The growing interest in different entropy functions towards black hole thermodynamics
as well as towards cosmology \cite{Li:2004rb,Li:2011sd,Wang:2016och,Nojiri:2005pu,Landim:2022jgr,Zhang:2005yz,Elizalde:2005ju,
Ito:2004qi,Gong:2004cb,Khurshudyan:2016gmb,Landim:2015hqa,Ghaffari:2019mrp,Li:2008zq,Zhang:2005hs,Li:2009bn,Feng:2007wn,Lu:2009iv,Nojiri:2017opc,Saridakis:2020zol,Barrow:2020kug,Lymperis:2023prf,Lymperis:2021qty,Nojiri:2019kkp,Paul:2019hys,Komatsu:2023wml,Luciano:2022viz,Lambiase:2023ryq,Nojiri:2023wzz} leads to the proposal of generalized entropy, depending on  number of parameters, which generalizes all the known and apparently different entropies (like the Bekenstein--Hawking entropy \cite{Bekenstein:1973ur,Hawking:1975vcx}, the Tsallis entropy \cite{Tsallis:1987eu}, the R\'{e}nyi entropy \cite{Renyi}, the Barrow entropy \cite{Barrow:2020tzx},  the Sharma--Mittal entropy \cite{SayahianJahromi:2018irq}, the Kaniadakis entropy \cite{Kaniadakis:2005zk} and the Loop Quantum gravity entropy 
 \cite{Majhi:2017zao}) for a suitable regime of the parameters \cite{Nojiri:2022aof,Nojiri:2022dkr,Odintsov:2022qnn,Odintsov:2023qfj}. Such interest in entropic cosmology becomes stronger when the entropic dark energy seems to be equivalent to holographic dark energy with suitable holographic cut-offs \cite{Nojiri:2021iko}. Initially, a six-parameter dependent generalized entropy of the form
\begin{eqnarray}
 S_\mathrm{6}(\alpha_\pm,\beta_\pm,\gamma_\pm) = \frac{1}{\alpha_++\alpha_-}\left[\left(1+\frac{\alpha_+}{\beta_+}S^{\gamma_+}\right)^{\beta_+}-\left(1+\frac{\alpha_-}{\beta_-}S^{\gamma_-}\right)^{-\beta_-}\right] \ ,
 \label{intro-1}
\end{eqnarray}
was proposed in 
 \cite{Nojiri:2022aof}, where $S = A/(4G)$ is the Bekenstein--Hawking entropy (with $A$ being the area of the apparent horizon and $G$ is the Newton's gravitational constant) and $\left\{\alpha_{\pm},\beta_{\pm},\gamma_{\pm}\right\}$ are the parameters. However, soon after \cite{Nojiri:2022aof}, a conjecture was made in \cite{Nojiri:2022dkr}, which stated that the minimum number of parameters required in a generalized entropy function that can generalize all the aforementioned entropies is equal to four. In particular, the four-parameter generalization is given by
\begin{eqnarray}
 S_\mathrm{4}(\alpha_\pm,\beta,\gamma) = \frac{1}{\gamma}\left[\left(1+\frac{\alpha_+}{\beta}S\right)^\beta-\left(1+\frac{\alpha_-}{\beta}S\right)^{-\beta}\right] \ ,
 \label{intro-2}
\end{eqnarray}
where $\left\{\alpha_{\pm},\beta,\gamma\right\}$ are the parameters which are considered to be positive in order to make $S_\mathrm{4}$ like a monotonic increasing function with respect to $S$. As a supporting argument of the conjecture, a counter example was shown in \cite{Nojiri:2022dkr} by an entropy function containing less than four
parameters (having three parameters, in particular) of the form
\begin{eqnarray}
 S_\mathrm{3}(\alpha,\beta,\gamma) = \frac{1}{\gamma}\left[\left(1+\frac{\alpha}{\beta}S\right)^\beta - 1\right]\ ,
 \label{intro-3}
\end{eqnarray}
which is not able to generalize all the known entropies; particularly, $S_\mathrm{3}$ does not represent Kaniadakis entropy in any situation. All the above entropies $\left\{S_\mathrm{6},S_\mathrm{4},S_\mathrm{3}\right\}$ possesses a singularity in a different type of cosmological scenario, particularly in a bouncing context. Such diverging behaviour is common to all the known entropies (like the Tsallis entropy, the R\'{e}nyi entropy, the Barrow entropy, the Sharma--Mittal entropy, the  Kaniadakis entropy and the Loop Quantum gravity entropy) as well as the Bekenstein--Hawking entropy itself diverges in a bouncing scenario (at the instant of bounce). A possible explanation of this issue is given in \cite{Odintsov:2022qnn}, where the authors proposed a singular-free generalized entropy containing five parameters of the form
\begin{eqnarray}
 S_\mathrm{5}(\alpha_\pm,\beta,\gamma,\epsilon) = \frac{1}{\gamma}\left[\left\{1+\frac{1}{\epsilon}\tanh{\left(\frac{\epsilon\alpha_+}{\beta}S\right)}\right\}^\beta-\left\{1+\frac{1}{\epsilon}\tanh{\left(\frac{\epsilon\alpha_-}{\beta}S\right)}\right\}^{-\beta}\right]\ ,
 \label{intro-4}
\end{eqnarray}
which is singular-free during the entire cosmological evolution of the universe even at a bouncing instant (in the context of bounce cosmology) and is able to generalize all the entropies known so far. According to the conjecture stated in \cite{Odintsov:2022qnn}, the minimum number of parameters required in a non-singular generalized entropy function that is able to generalize all the previously known entropies is equal to five. Based on universe's evolution, in particular, whether the universe passes through a non-singular bounce (or not) during its cosmic evolution, the minimal constructions of generalized version of entropy is given by the four-parameter \cite{Nojiri:2022dkr} and the five-parameter \cite{Odintsov:2022qnn} generalized entropy, respectively. Various representatives of $\left\{S_\mathrm{6},S_\mathrm{4},S_\mathrm{3},S_\mathrm{5}\right\}$ and their convergence to the known entropies are schematically shown in Table~\ref{table}. The wide applications of the generalized entropies towards cosmology as well as towards black holes are addressed in \cite{Nojiri:2022dkr,Odintsov:2023qfj,Nojiri:2022nmu,Odintsov:2023vpj,Nojiri:2022ljp,Bolotin:2023wiw,Odintsov:2023weg}. Here, it also deserves  mention that the microscopic interpretation of such generalized entropies were not known until recently, when some of our authors gave a statistical description of the same in microcanonical, canonical and grand-canonical ensemble \cite{Nojiri:2023ikl,Nojiri:2023bom}. 

Based on the above arguments, we will work with  four-parameter generalized entropy in the present work, which contains the minimum number of parameters and also generalizes all the known entropies so far. In particular, we will concentrate on early universe cosmology with four-parameter generalized entropy, where the matter fields inside of the apparent horizon are dominated by a minimally coupled scalar field with a power law type potential. Actually, without the matter fields inside the horizon, the entropic cosmology results in a de-Sitter spacetime, or equivalently, an eternal inflation having no exit. Thus, in order to have a viable inflation, one needs to take either of the following approaches---(a) the entropic parameters vary with the cosmic expansion of the universe, or (b) by some matter fields inside the horizon. In the context of generalized entropy, the first possibility has been studied in \cite{Nojiri:2022dkr}, while the second approach will be examined in the present work where the matter fields are taken to be a scalar field with a power law potential. Such a form of  scalar potential is motivated by the fact that the simplest $\phi^2$ potential (where $\phi$ is the scalar field under consideration) in standard scalar field cosmology is ruled out from inflationary perspective as it is not consistent with the recent Planck 2018 data; (note that this is particular case of FRW cosmology \cite{FRW1}, actually in the realm of ``apparent horizon thermodynamics'', the standard scalar field cosmology is analogous to the case where the entropy of the apparent horizon is given by the Bekenstein--Hawking entropy, which produces the usual Friedmann equations from the thermodynamic law of the apparent horizon). However the story becomes different if the horizon entropy is of the generalized entropic form, in which case, the effective energy density coming from the horizon entropy plays a significant role during the evolution of the universe. Motivated by this, we intend to examine the status of the $\phi^2$ potential in the context of entropic inflation where the horizon entropy is given by the four-parameter generalized entropy.

\begin{table}[H]
\caption{\label{table}Schematic table to summarize various representatives of the generalized entropies and their convergence to the known entropies. 
 Here, $S_\mathrm{T} = \textrm{Tsallis entropy}$, \mbox{$S_\mathrm{B} = \textrm{Barrow entropy}$}, $S_\mathrm{R} = \textrm{R\'{e}nyi entropy}$, $S_\mathrm{SM} = \textrm{Sharma--Mittal entropy}$, $S_\mathrm{K} = \textrm{Kaniadakis entropy}$ and \mbox{$S_\mathrm{q} = \textrm{Loop Quantum gravity entropy}$}.}

\setlength{\tabcolsep}{30pt} 
\renewcommand{\arraystretch}{1.1} 

\begin{tabularx}{\textwidth}{ccc}
\toprule
 \multirow{5}{*}{\Large$S_\mathrm{3}$}
& $\gamma=\alpha$ &$S_\mathrm{SM}$  \\
    & $\alpha\rightarrow\infty$ & $S_\mathrm{T},S_\mathrm{B}$ \\
    & $\alpha,\beta\rightarrow 0$ with $\frac{\alpha}{\beta}$ finite & $S_\mathrm{R}$\\
    & $\beta \rightarrow \infty, \gamma=\alpha$ & $S_\mathrm{q}$ 
     \\\midrule
 \multirow{5}{*}{\Large$S_\mathrm{5}$}
& $\epsilon,\alpha_-\rightarrow0,\alpha_+=\gamma$ &$S_\mathrm{SM}$  \\
    & $\epsilon\rightarrow0,\alpha_-=0,\alpha_+\rightarrow\infty,\gamma=\left(\frac{\alpha_+}{\beta}\right)^\beta$ & $S_\mathrm{T},S_\mathrm{B}$ \\
    & $\epsilon,\beta\rightarrow0,\alpha_-=0,\alpha_+=\gamma$ with $\frac{\alpha_+}{\beta}$ finite & $S_\mathrm{R}$\\
    & $\epsilon,\alpha_-\rightarrow0,\beta\rightarrow\infty,\alpha_+=\gamma$ & $S_\mathrm{q}$  \\
    &$\epsilon\rightarrow0,\beta\rightarrow\infty,\alpha_+=\alpha_-$ & $S_\mathrm{K}$  \\\midrule

 \multirow{5}{*}{\Large$S_\mathrm{4}$}
& $\alpha_-=0,\alpha_+=\gamma$ &$S_\mathrm{SM}$  \\
    & $\alpha_+\rightarrow\infty,\alpha_-=0$ & $S_\mathrm{T},S_\mathrm{B}$ \\
    & $\alpha_-=0,\alpha_+=\gamma,\beta\rightarrow0$ with $\frac{\alpha_+}{\beta}$ finite & $S_\mathrm{R}$\\
    & $\beta\rightarrow\infty,\alpha_-=0,\alpha_+=\gamma$ & $S_\mathrm{q}$  \\
    & $\beta \rightarrow \infty, \alpha_+=\alpha_-$ & $S_\mathrm{K}$  \\\midrule
     \multirow{5}{*}{\Large$S_\mathrm{6}$}
& $\alpha_-=0,\alpha_+=\gamma_+\beta_+$ &$S_\mathrm{SM}$  \\
    & $\alpha_+=\alpha_-\rightarrow0,\gamma_+=\gamma_-$ & $S_\mathrm{T},S_\mathrm{B}$ \\
    & $\alpha_+,\beta_+\rightarrow0,\gamma_+=1$ with $\frac{\alpha_+}{\beta_+}$ finite & $S_\mathrm{R}$\\
    & $\beta_+\rightarrow\infty,\alpha_-=0,\gamma_+=1$ & $S_\mathrm{q}$  \\
    & $\beta_\pm \rightarrow 0, \alpha_+=\alpha_-,\gamma_\pm=1$ & $S_\mathrm{K}$  \\ \bottomrule
\end{tabularx}

\end{table}

The paper is organized as follows: the modified Friedmann equations for four-parameter generalized entropy is discussed in Section~\ref{sec-modified-cosmology}. Then after giving a brief review of $\phi^n$ inflationary potential with the Bekenstein--Hawking entropy (i.e in standard scalar field cosmology) in Section~\ref{sec-BH-SF}, we will examine the status of the same inflationary potential with 4-parameter generalized entropy as the horizon entropy in Section~\ref{sec-gen-SF}. The paper ends with some conclusions in Section~\ref{sec-conclusion}.

\section{Modified Cosmology with Generalized Entropy}\label{sec-modified-cosmology}

We consider a spatially flat and isotropic universe described by the Friedmann--Lema\^{i}tre--Robertson--Walker (FRLW) metric
\begin{equation}
    ds^2 = - dt^2 + a(t) \Big(dr^2 + r^2 \left(d\theta^2+\sin^2{\theta} \, d\phi^2\right)\Big) \ .
\end{equation}
It can be rewritten in the following way
\begin{equation}
    ds^2 = h_{ab}dx^a dx^b + \Tilde{r}^2 \left(d\theta^2+\sin^2{\theta} \, d\phi^2\right) \ ,
\end{equation}
defining $\Tilde{r} (r,t) = a(t) r$, $h_{ab} = \text{diag}(-1, a^2)$
and $x^0=t$ , $x^1=r$. The apparent horizon is defined by $h^{ab}\partial_a \Tilde{r}\partial_b \Tilde{r} =0$, which in the case of a spatially flat FLRW
background has the solution \cite{Cai:2005ra,Akbar:2006kj,Sanchez:2022xfh}
\begin{equation}\label{RadHorAppFlat}
    r_h = \frac{1}{H} \ .
\end{equation}
In this background we see that the apparent horizon is equivalent to the cosmological horizon, that is the Hubble radius. Consequently, we can define a temperature $T= |\kappa|/(2\pi)$, where $\kappa$ is the surface gravity defined by \cite{Cai:2005ra}
\begin{equation}
    \kappa = \frac{1}{2\sqrt{-h}}\partial_a\left(\sqrt{-h}h^{ab}\partial_b\Tilde{r}\right)\Big|_{\Tilde{r}=r_h} \ ,
\end{equation}
which can be rewritten as
\begin{equation}
    \kappa = - \frac{1}{r_h}\left(1-\frac{\dot{r}_h}{2 H r_h}\right) \ ,
\end{equation}
and it leads to a temperature
\begin{equation}
    T = \frac{1}{2\pi r_h}\left|1 -\frac{\dot{r}_h}{2 H r_h}\right|\ .
\end{equation}
As in the case of the \BH 
entropy we can then associate a generalized entropy $S_g$ to the apparent horizon in order to find the field equations. The first principle of thermodynamics states \cite{Akbar:2006kj,Sanchez:2022xfh}
\begin{equation}
    T dS_g = - dE + W dV \ ,
\end{equation}
where $V$ is the volume of the apparent horizon, $E = \rho V$ is the total internal energy inside of the horizon and $W = \frac{1}{2}\left(\rho - p\right)$ represents the work density regarding the thermodynamic law. The right side of this equation takes the expression as follows:
\begin{equation}
    T dS_g = - V d\rho -\frac{1}{2}(\rho +p) dV \ .
\end{equation}
To express this differential equation in terms of the apparent horizon we rewrite
\begin{equation}
    T dS_g = T \frac{\partial S_g}{\partial S} dS = \frac{1}{G}\left|1-\frac{\Dot{r}_h}{2 H r_h}\right| \frac{\partial S_g}{\partial S} d r_h
\end{equation}
and
\begin{equation}
    - V d\rho -\frac{1}{2}(\rho + p) dV = -\frac{4 \pi}{3} r_h^3 \left( d\rho -\dot{\rho}\frac{dr_h}{2H r_h}\right) \ ,
\end{equation}
where we used the conservation equation of the matter fields $\Dot{\rho} + 3 H(\rho+p)=0$. At this point equaling the two terms we can derive the field equation for a general dynamical apparent horizon $r_h$
\begin{equation}
   \frac{\partial S_g}{\partial S} \frac{\Dot{r}_h}{r_h^3} = -\frac{4 \pi G}{3} \Dot{\rho} \ ,
\end{equation}
which, for the choice of $r_h = 1/H$ along with the conservation relation of matter fields, becomes
\begin{eqnarray}
 \Dot{H}\left(\frac{\partial S_g}{\partial S}\right) = -4\pi G\left(\rho + p\right) \ ,
 \label{FriedEQGen-2}
\end{eqnarray}
which is considered to be the second Friedmann equation in the context of generalized entropic cosmology.
The integration of such equation leads to the first Friedmann equation corresponding to the generalized entropy we are considering as
\begin{equation}\label{FriedEQGen}
    \int d\left(H^2\right) \left(\frac{\partial S_g}{\partial S}\right) = \frac{8 \pi G}{3} \rho + \frac{\Lambda}{3} \ ,
\end{equation}
where $\Lambda$, known as the cosmological constant, appears as an integration constant. The above two equations represent the general Friedmann equations based on the apparent horizon thermodynamics for any form of horizon entropy.

For the three-parameter and the four-parameter generalized entropy, Equation~(\ref{FriedEQGen-2}) takes the following form:
\begin{eqnarray}
 \Dot{H}\left\{\frac{\alpha}{\gamma}\left(1+\frac{\alpha}{\beta}S\right)^{\beta-1}\right\} = -4\pi G\left(\rho + p\right) \ ,
 \label{II_Fried_S3}
\end{eqnarray}
and
\begin{eqnarray}
 \Dot{H}\left\{\frac{1}{\gamma}\left[\alpha_+\left(1+\frac{\alpha_+}{\beta}S\right)^{\beta-1}+\alpha_-\left(1+\frac{\alpha_-}{\beta}S\right)^{-\beta-1}\right]\right\} = -4\pi G\left(\rho + p\right) \ ,
 \label{II_Fried_S4}
\end{eqnarray}
respectively. Moreover the first Friedmann equation, i.e., Equation~(\ref{FriedEQGen}), yields
\begin{equation}\label{I_Fried_S3}
    \frac{\beta GH^4}{\pi\gamma(2-\beta)}\left(\frac{\beta GH^2}{\pi\alpha}\right)^{-\beta} 2F_1\left(1-\beta,2-\beta,3-\beta,-\frac{\beta G H^2}{\pi\alpha}\right) = \frac{8 \pi G}{3} \rho + \frac{\Lambda}{3} \ ,
\end{equation}
for $S_\mathrm{3}$, and
\begin{align} \label{I_Fried_S4}
    \frac{\beta G H^4}{\pi\gamma}\Bigg[ &\frac{1}{2+\beta}\left(\frac{\beta GH^2}{\pi\alpha_-}\right)^{\beta} 2F_1\left(1+\beta,2+\beta,3+\beta,-\frac{\beta G H^2}{\pi\alpha_-}\right)  \nonumber\\& +\frac{1}{2-\beta}\left(\frac{\beta GH^2}{\pi\alpha_+}\right)^{-\beta} 2F_1\left(1-\beta,2-\beta,3-\beta,-\frac{\beta G H^2}{\pi\alpha_+}\right)\Bigg]= \frac{8 \pi G}{3} \rho + \frac{\Lambda}{3} \ ,
\end{align}
for $S_\mathrm{4}$; where $2F_1~(\textrm{arguments})$ symbolizes the Hypergeometric function. Owing to the above equations, we may argue that the generalized entropy generates an effective energy density (along with the normal matter fields) in the Friedmann equation; for instance, the energy density coming from the $S_\mathrm{4}$ is given by,
\begin{eqnarray}
 \rho_\mathrm{g} = \frac{3}{8\pi G}\Bigg\{H^2&-&\frac{\beta G H^4}{\pi\gamma}\Bigg[\frac{1}{2+\beta}\left(\frac{\beta GH^2}{\pi\alpha_-}\right)^{\beta} 2F_1\left(1+\beta,2+\beta,3+\beta,-\frac{\beta G H^2}{\pi\alpha_-}\right)\nonumber\\
 &+&\frac{1}{2-\beta}\left(\frac{\beta GH^2}{\pi\alpha_+}\right)^{-\beta} 2F_1\left(1-\beta,2-\beta,3-\beta,-\frac{\beta G H^2}{\pi\alpha_+}\right)\Bigg]\Bigg\} \ ,
 \nonumber
\end{eqnarray}
and consequently, Equation~(\ref{I_Fried_S4}) can be written as,
\begin{eqnarray}
 H^2 = \frac{8\pi G}{3}\left(\rho + \rho_\mathrm{g}\right) + \frac{\Lambda}{3} \ .
 \nonumber
\end{eqnarray}
Similarly the energy density corresponding to the  3-parameter generalized entropy can be determined from Equation~(\ref{I_Fried_S3}).

The energy density coming from the generalized entropy plays a significant role during the evolutionary course of the universe. However, without any matter fields inside of the horizon, Equation~(\ref{II_Fried_S4}) (or Equation~(\ref{II_Fried_S3})) shows $\Dot{H} = 0$ leading to $H = \mathrm{constant}$. This argues that the entropic cosmology, in absence of matter fields, results in a de-Sitter spacetime, or equivalently, an eternal inflation having no exit. Therefore, in order to obtain a viable inflation, one needs to incorporate either of the following possibilities---(a) the entropic parameters slowly vary with the cosmic expansion of the universe {(one may see~\cite{DiGennaro:2022ykp} where the authors studied a energy scale-varying entropic index that could lead to new physics in the early universe)}, or (b) some matter fields inside of the horizon. In the present work, we will concentrate on the second possibility, where the matter field is taken to be a minimally coupled scalar field with a power law potential. Such form of the scalar potential in the context of generalized entropy is well motivated, as discussed in the introduction. However before moving to the case of the generalized entropy of the apparent horizon, we will discuss the status of $\phi^n$ potential (from inflationary perspective) with the Bekenstein--Hawking entropy in order to understand the role of the generalized entropy during the early evolution of the universe. These are the subjects of the next sections. (Moreover, the cases with varying entropic parameters, in the context of three-parameter and four-parameter generalized entropies, are addressed in the Appendix, see Appendix~\ref{sec-appendix}).


\section{Status of $\phi^n$ Inflationary Potential with Bekenstein--Hawking Entropy}\label{sec-BH-SF}
In this section we will investigate whether a $\phi^n$ type of potential, in the case where the entropy of the apparent horizon is given by the Bekenstein--Hawking entropy, can lead to a viable inflation during the early universe. As a result, the Equations~(\ref{FriedEQGen-2}) and~(\ref{FriedEQGen})  read as (by considering $S_\mathrm{g} = S$)
\begin{align}\label{N1}
    H^2 &= \frac{8 \pi G}{3}\left\{\frac{\Dot{\phi}^2}{2}+V(\phi)\right\} \\
    \Dot{H} &= -4\pi G \Dot{\phi}^2 \ ,
\end{align}
respectively, and the continuity equation for the scalar field becomes
\begin{equation}\label{N2}
    \Ddot{\phi}+3 H \Dot{\phi}+\partial_\phi V = 0 \ .
\end{equation}
The above equations are similar to that of in the standard scalar field cosmology---this is however expected, as the Bekenstein--Hawking entropy leads to the usual Friedmann equations in the realm of entropic cosmology. By the slow roll approximation, i.e., by assuming that the potential energy during inflation dominates all the other forms of energies, the first and second Friedmann equations become
\begin{align}
    H^2 = \frac{8 \pi G}{3} V(\phi)~~~~~~~~~~~~~\mathrm{and}~~~~~~~~~~~~~~\Dot{H} = -4\pi G \Dot{\phi}^2\ ,
\end{align}
respectively, and moreover, the continuity equation is approximated as,
\begin{equation}\label{SR-continuity}
    \Dot{\phi} = -\left(\frac{1}{3H}\right)\partial_\phi V = 0 \ .
\end{equation}
Consequently the slow roll parameters take the following form,
\begin{align}\label{epsetaStandard}
    \epsilon(t) = \frac{3}{2}\frac{\Dot{\phi}^2}{V(\phi)}~~~~~~\textrm{and}~~~~~~~\eta(t) = -\sqrt{\frac{3}{8\pi G} \frac{1}{V(\phi)}}\frac{\Ddot{\phi}}{\Dot{\phi}} \ .
\end{align}
For the scalar potential to be of the form like $V(\phi) = V_0 \phi^n$, the dynamical equation for the scalar field from Equation~(\ref{SR-continuity}) obtains the expression
\begin{equation}
    \Dot{\phi} \simeq -\frac{1}{3 H} \partial_\phi V = - \frac{n V_0}{3}\left(\frac{8\pi G}{2} V_0\right)^{-\frac{1}{2}}\phi^{\frac{n}{2}-1} \ ,
\end{equation}
by using which, into Equation~(\ref{epsetaStandard}), we obtain the slow roll parameters in terms of scalar field as follows:
\begin{align}\label{epsetaStandard-1}
    \epsilon(\phi) = \frac{n^2}{16 \pi G} \phi^{-2}~~~~~~~~\textrm{and}~~~~~~~~\eta (\phi) = \frac{n(n-2)}{16\pi G}\phi^{-2} \ ,
\end{align}
respectively, where both $\epsilon(\phi)$ and $\eta(\phi)$ are inversely proportional to $\phi$, i.e., both of the slow roll parameters increases with the decreasing value of the scalar field. Such behaviour of $\epsilon(\phi)$ actually helps to trigger a viable inflation. In particular, a considerably large value of $\phi$ makes $\epsilon(\phi)$ less than unity, which confirms an accelerated stage of the universe. However as the scalar field rolls down (from a larger value to a smaller value, governing by Equation~(\ref{SR-continuity})), $\epsilon(\phi)$ increases, and at a certain instance of time, $\epsilon(\phi)$ becomes unity, which indicates the end of inflation. Let us consider that $\epsilon(\phi)$ becomes unity at $\phi = \phi_\mathrm{f}$, i.e., $\epsilon(\phi_\mathrm{f}) = 1$, which, along with Equation~(\ref{epsetaStandard-1}), yields the following form of $\phi_\mathrm{f}$ (in terms of model parameters):
\begin{equation}
    \phi_f^2 = \frac{n^2}{16 \pi G} \ .
\end{equation}
The total number of e-folds of inflationary epoch is given by
\begin{equation}\label{e-fold}
    N_\mathrm{f} = \int_{\phi_c}^{\phi_f} \frac{H}{\Dot{\phi}} d\phi \simeq - \int_{\phi_c}^{\phi_f} \frac{3H^2}{\partial_\phi V} d\phi = -\frac{4 \pi G }{n} \left(\phi_f^2-\phi_c^2\right) = \frac{4 \pi G }{n} \left(\phi_c^2-\frac{n^2}{16 \pi G}\right) \ ,
\end{equation}
where in the last step we used the condition $\epsilon(\phi_f)=1$. Here, $\phi_\mathrm{c}$ is the scalar field at the time of horizon crossing of the CMB mode ($\sim 0.05\mathrm{Mpc}^{-1}$ at which we are interested to determine the observable parameters). Inverting Equation~(\ref{e-fold}), we immediately obtain $\phi_\mathrm{c}$ in terms of $N_\mathrm{f}$ as,
\begin{equation}
    \phi_\mathrm{c} = \sqrt{\frac{n}{16\pi G}\left(4N_\mathrm{f} + n\right)} \ ,
\end{equation}
so that we can then compute the slow roll parameters at the instant of horizon crossing of the CMB scale modes, and they are given by,
\begin{align}\label{Standardnsr}
    \epsilon(\phi_c) = \frac{n}{4N_\mathrm{f} + n}~~~~~~~~~~~\textrm{and}~~~~~~~~~~~~\eta (\phi_c) = \frac{n-2}{4N_\mathrm{f} + n} \ .
\end{align}
These will be used to compute the spectral tilt for primordial curvature perturbation ($n_s$) and the tensor-to-scalar ratio ($r$) that are defined by
\begin{eqnarray}
n_s = 1-6\epsilon+2\eta\bigg|_{\phi=\phi_c}~~~~~~~~~~~~~\mathrm{and}~~~~~~~~~~~~~~r = 16\epsilon\bigg|_{\phi=\phi_c} \ ,
\end{eqnarray}
respectively, at the horizon crossing instant. Using the expressions of $\epsilon(\phi_c)$ and $\eta (\phi_c)$ into the bove equation along with a little bit of simplification yields the final forms of $n_s$ and $r$ as follows:
\begin{eqnarray}\label{ns-r-final}
 n_s = 1 - \frac{4(n+1)}{4N_\mathrm{f} + n}~~~~~~~~~~~~~~~\textrm{and}~~~~~~~~~~~~~~r = \frac{16n}{4N_\mathrm{f} + n} \ .
\end{eqnarray}
Having Equation~(\ref{ns-r-final}) in hand, we now examine the status of $\phi^n$ potential with the Bekenstein--Hawking entropy of the apparent horizon, in respect to the Planck 2018 data which puts a constraint on the observable indices as \cite{Planck:2018jri}:
\begin{equation}\label{constraint-ns-r}
    n_s = 0.9649 \pm 0.0042\quad \text{and} \quad r < 0.064 \ .
\end{equation}
It is evident from Equation~(\ref{ns-r-final}) that both the $n_s$ and $r$ depend on $n$ and $N_\mathrm{f}$. In Figure~\ref{fig:StandardRegion} we plot the region of validity for the inflationary indices in the case of the Bekenstein--Hawking entropy in the $n\textrm{-}N_\mathrm{f}$ space (we will concentrate around $N_\mathrm{f} = 60$, which is consistent with the resolution of the horizon problem). It may be noted that we consider $n\geq 1$ in the plot, as $n<1$ generates some singularity problem in the scalar field equation (through $\frac{\partial V}{\partial\phi}$) when the scalar field passes through $\phi = 0$. We see from this plot that the two regions barely overlaps in a small region (near at $n=1$) far from the expected value of $n_s$, which can be considered statistically negligible. Therefore, in the scenario where the horizon entropy is given by the Bekenstein--Hawking entropy and the scalar field inside the horizon has a $\phi^n$ form of potential, there is no choice of the parameter $n$ that provides the simultaneous agreement of $\left\{n_s,r\right\}$ with the Planck observation.

\begin{figure}[!h]

\includegraphics[width=3.5in,height=2.5in]{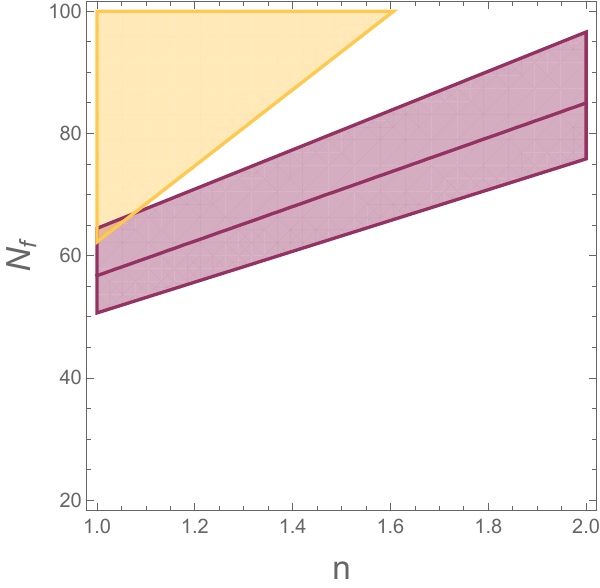}
\caption{Region of validity for the observable indices $n_s$ (Purple) and $r$ (Orange) given in \mbox{Equation~(\ref{ns-r-final})} in the $n\textrm{-}N_\mathrm{f}$ space. The purple region corresponds to the constraints region of $n_s$ while the purple line represents its central value.}
    \label{fig:StandardRegion}
 \label{fig:StandardRegion}

\end{figure}


\section{Status of $\phi^n$ Inflationary Potential with Generalized Entropy}\label{sec-gen-SF}

As showed in Section~[\ref{sec-BH-SF}] that the simple $V(\phi)$$\sim$$\phi^2$ potential with the Bekenstein--Hawking entropy for the apparent horizon does not lead to a viable inflation from the perspective of the Planck 2018 data. However the story of the $\phi^2$ inflaton potential may become different in generalized entropic cosmology, in which case, the entropic energy density arising from the generalized entropy contributes a significant role during the universe's evolution. Thus we will investigate the status of $V(\phi)$$\sim$$\phi^2$ potential in the context of generalized entropic cosmology, where the presence of entropic energy density results in a different cosmological scenario compared to that of in the Bekenstein--Hawking entropic scenario. In particular, we will consider that the horizon entropy is of the form of four-parameter generalized entropy and the matter fields inside of the horizon is dominated by a scalar field having $V(\phi)$$\sim$$\phi^{n}$ potential.

\begin{figure}[H]

    \includegraphics[scale=0.8]{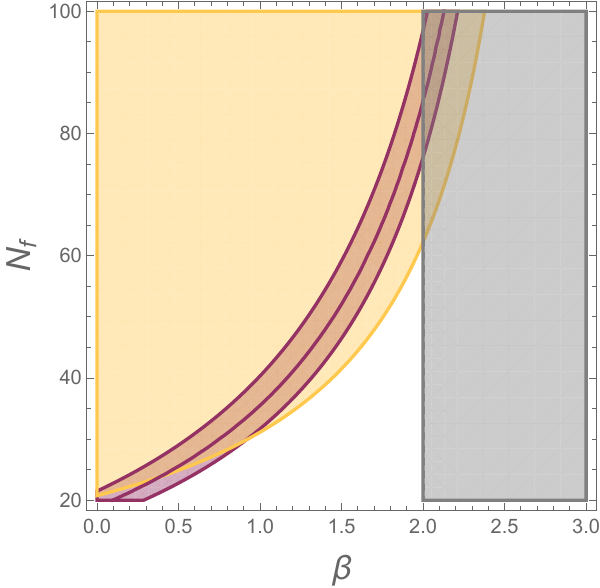}
    \caption{Region of validity of the observable indices 
 $n_s$ (purple) and $r$ (orange) given in Equation~(\ref{ns-r-gen}) in respect to the Planck data. Here we take $n=1$. The gray shadowed region, corresponding to $\beta > 2$, is not acceptable since it leads to negative values of $\epsilon$. The purple region corresponds to the allowed region of $n_s$ while the purple line represents its central value (i.e., $n_s= 0.9649$).}
    \label{fig:3ParRegionn1}
\end{figure}

\begin{figure}[H]
   \includegraphics[scale=0.8]{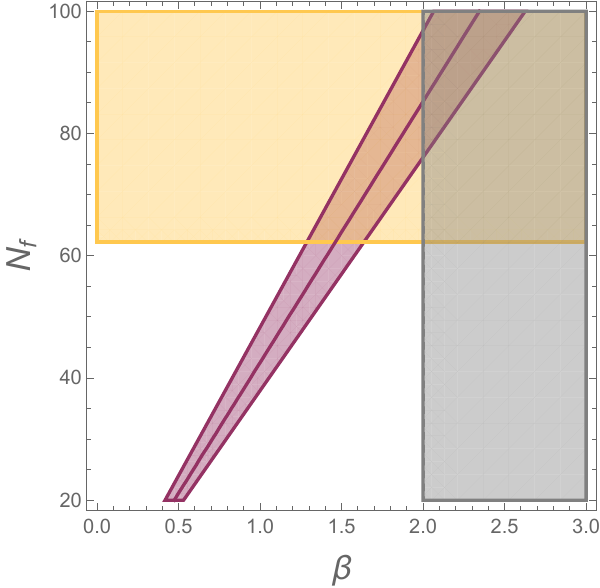}
    \caption{{Region of validity of the observable indices}
 $n_s$ (purple) and $r$ (orange) given in Equation~(\ref{ns-r-gen}) in respect to the Planck data. Here we take $n=2$. The gray shadowed region, corresponding to $\beta > 2$, is not acceptable since it leads to negative values of $\epsilon$. The purple region corresponds to the allowed region of $n_s$ while the purple line represents its central value (i.e., $n_s= 0.9649$).}
    \label{fig:3ParRegionn2}

\end{figure}

For the case of four-parameter generalized entropy, the first Friedmann Equation~(\ref{I_Fried_S4}), without any cosmological constant, is given by:
\begin{align}\label{I-FRW-S4-2}
\frac{\beta G H^4}{\pi\gamma}\Bigg[ &\frac{1}{2+\beta}\left(\frac{\beta GH^2}{\pi\alpha_-}\right)^{\beta} 2F_1\left(1+\beta,2+\beta,3+\beta,-\frac{\beta G H^2}{\pi\alpha_-}\right)  \nonumber\\& +\frac{1}{2-\beta}\left(\frac{\beta GH^2}{\pi\alpha_+}\right)^{-\beta} 2F_1\left(1-\beta,2-\beta,3-\beta,-\frac{\beta G H^2}{\pi\alpha_+}\right)\Bigg]= \frac{8 \pi G}{3} \rho \ .
\end{align}
The cosmological constant during the early phase of the universe is suppressed compared to the inflaton energy density and, thus, one can safely neglect the $\Lambda$ in studying the inflationary dynamics of the universe. Since during the inflation the typical energy scale is of order $\sim 10^{-4}M_{Pl}$ (where $M_\mathrm{Pl} = 1/\sqrt{8\pi G}$ with $G$ being the Newton's gravitational constant), we consider $G H^2 \ll 1$ during the early universe. As a consequence, we can expand the hypergeometric functions appearing in Equation~(\ref{I-FRW-S4-2}), thanks to the relation $2F_1(a,b,c,x) = 1+\frac{ab}{c}x+O(x^2)$, leading to
\begin{eqnarray}\label{FriedEqComp}
\frac{1}{2-\beta}\frac{\beta G H^4}{\pi\gamma}\left(\frac{\beta GH^2}{\pi\alpha_+}\right)^{-\beta} \left(1-\frac{(1-\beta)(2-\beta)}{3-\beta}\frac{\beta G H^2}{\pi\alpha_+}\right)= \frac{8 \pi G}{3} \rho \ ,
\end{eqnarray}
on solving which, at the leading order in $GH^2$, we obtain
\begin{equation}\label{FriedI_inf}
    H^2 = \left( \frac{8\pi G}{3}\frac{\gamma}{\alpha_+}\left(\frac{\beta G }{\pi \alpha_+}\right)^{\beta-1}(2-\beta)\, \rho \,\right)^{\frac{1}{2-\beta}} \ .
\end{equation}
Moreover the second Friedmann Equation~(\ref{FriedEQGen-2}), due to $GH^2 \ll 1$, takes the following form
\begin{equation}\label{FriedII_inf}
    \frac{\alpha_+}{\gamma}\left(\frac{\alpha_+}{\beta}\frac{\pi}{G H^2}\right)^{\beta-1}\Dot{H} = -4\pi G(\rho+p) \ .
\end{equation}
As mentioned earlier that we will consider a minimally coupled scalar field as the matter field inside of the horizon, for which, the corresponding energy density ($\rho$) and the pressure ($p$) are given by
\begin{align}
    \rho = \frac{\Dot{\phi}^2}{2}+V(\phi)~~~~~~~~~~~~~~\textrm{and}~~~~~~~~~~~~~~p = \frac{\Dot{\phi}^2}{2}-V(\phi) \ ,
\end{align}
respectively, where $\phi$ is the scalar field under consideration and $V(\phi)$ is its potential. Therefore, Equations~(\ref{FriedI_inf}) and~(\ref{FriedII_inf}) become
\begin{equation}\label{FriedI_inf-2}
    H^2 = \left( \frac{8\pi G}{3}\frac{\gamma}{\alpha_+}\left(\frac{\beta G }{\pi \alpha_+}\right)^{\beta-1}(2-\beta)\, \left\{\frac{\Dot{\phi}^2}{2}+V(\phi)\right\} \,\right)^{\frac{1}{2-\beta}} \ .
\end{equation}
and
\begin{equation}\label{FriedII_inf-2}
    \frac{\alpha_+}{\gamma}\left(\frac{\alpha_+}{\beta}\frac{\pi}{G H^2}\right)^{\beta-1}\Dot{H} = -4\pi G \Dot{\phi}^2 \ .
\end{equation}
The above two equations along with the continuity equation of the scalar field, i.e.,
\begin{equation} 
    \Ddot{\phi}+3 H \Dot{\phi}+\partial_\phi V = 0 \ .
\end{equation}
govern the cosmological dynamics in the present context. Due to the slow roll approximation during the early universe, in particular
\begin{equation}
    \Ddot{\phi}\ll H\Dot{\phi} \hspace{1cm} \text{and}\hspace{1cm} \frac{\Dot{\phi}^2}{2}\ll V \ ,
\end{equation}
Equations~(\ref{FriedI_inf-2}) and~(\ref{FriedII_inf-2}) read as
\begin{align}\label{SR-inflation-equation-1}
    H^2 &=\left[ \frac{8 \pi G}{3} \frac{\gamma}{\alpha_+}\left(\frac{\beta G}{\alpha_+ \pi}\right)^{\beta-1}(2-\beta)\, V(\phi) \,\right]^\frac{1}{2-\beta} \\
    \Dot{H} &= -\frac{2\pi G\gamma}{\alpha_+}\left(\frac{\beta G H^2}{\alpha_+\pi}\right)^{\beta-1}\Dot{\phi}^2 \ ,
\end{align}
and moreover, the conservation equation is approximated as
\begin{equation}\label{SR-inflation-equation-2}
    \Dot{\phi} \simeq - \frac{1}{3H}\partial_\phi V \ .
\end{equation}
Therefore the first and the second slow roll parameters (defined by $\epsilon(t) = -\frac{\Dot{H}}{H^2}$ and  $\eta(t) = -\frac{\Ddot{H}}{2 H \Dot{H}}$) take the following form
\begin{align}\label{SR-parameter-inf-1}
    \epsilon(t) &= \frac{3}{4(2-\beta)}\frac{\Dot{\phi}^2}{V(\phi)} \\ 
    \eta(t) &= -\left[ \frac{8 \pi G}{3} \frac{\gamma}{\alpha_+}\left(\frac{\beta G}{\alpha_+ \pi}\right)^{\beta-1}(2-\beta)\, V(\phi) \,\right]^\frac{1}{2(2-\beta)}\left[\frac{\Ddot{\phi}}{\Dot{\phi}}+\frac{1-\beta}{4(2-\beta)}\frac{\partial_\phi V }{V(\phi)}\Dot{\phi}\right] \ ,
\end{align}
where for the second parameter we have used Equation~(\ref{SR-inflation-equation-2}). At this stage, let us consider the power law form of the scalar potential, i.e., $V(\phi) = V_0\phi^n$ (with $V_0$ and $n$ being two positive constants). For this scalar potential, we can compute the expression of $\Dot{\phi}$ from Equation~(\ref{SR-inflation-equation-2}), and is given by,
\begin{equation}
    \Dot{\phi} \simeq -\frac{V_0}{3} n \, A^{-\frac{1}{2(2-\beta)}} \,  \phi^{\frac{3-2\beta}{2(2-\beta)}n-1} \ ,
\end{equation}
where we have
\begin{equation}\label{A}
    A \equiv  \frac{8 \pi G}{3} \frac{\gamma}{\alpha_+}\left(\frac{\beta G}{\alpha_+ \pi}\right)^{\beta-1}(2-\beta)\, V_0 \ .
\end{equation}
Plugging back the above expression of $\Dot{\phi}$ into Equation~(\ref{SR-parameter-inf-1}) yields the slow roll parameters in terms of $\phi$ as follows:
\begin{align} \label{InfParPhi}
    \epsilon(\phi) &=\frac{V_0 n^2}{12(2-\beta)} A^{-\frac{1}{2-\beta}} \, \phi^{\frac{1-\beta}{2-\beta}n-2}\\ 
    \eta(\phi) &= \frac{V_0}{2}n \left(\frac{7-5\beta}{4(2-\beta)}n-1\right) \, A^{-\frac{1}{2-\beta}}\, \phi^{\frac{1-\beta}{2-\beta}n-2} \ ,
\end{align}
where $A$ is given above in Equation~(\ref{A}). It seems that the positivity of $\epsilon$ demands $\beta < 2$. Moreover the above expression of $\epsilon(\phi)$ clearly points that the model parameters ($\beta$ and $n$) need to obey the following constraint relation, namely
\begin{eqnarray}\label{constraint}
 \left(\frac{1-\beta}{2-\beta}\right)n < 2
\end{eqnarray}
in order to have a successful inflation with an exit. This is because that under  \mbox{Condition~(\ref{constraint})}, $\epsilon(\phi)$ remains less than unity for a considerably large positive value of $\phi$ and triggers an accelerating stage of the universe; however, as the scalar field rolls down along the potential, $\epsilon(\phi)$ increases and moves to unity at a certain value of $\phi = \phi_\mathrm{f}$ (say) which indicates the end of inflation. Thus as a whole---owing to Condition (\ref{constraint})---$\epsilon(\phi)$ depends on the inverse power of $\phi$, which proves to be suitable for obtaining a successful inflation with a graceful exit. The end point of inflation, i.e $\epsilon(\phi_\mathrm{f}) = 1$, immediately leads to $\phi_\mathrm{f}$ from Equation~(\ref{InfParPhi}) as follows:
\begin{equation}\label{ConditionEpsInf}
    A^\frac{1}{2-\beta}\phi_f^{2-\frac{1-\beta}{2-\beta}n} = \frac{V_0 n^2}{12(2-\beta)} \ .
\end{equation}
Consequently, the e-fold duration of the inflationary era is given by,
\begin{equation}
    \int_0^{N_\mathrm{f}} dN = \int_{t_\mathrm{c}}^{t_\mathrm{f}} H dt = \int_{\phi_\mathrm{c}}^{\phi_\mathrm{f}} \frac{H}{\Dot{\phi}} d\phi \ ,
\end{equation}
from which, using the conservation equation, we obtain
\begin{equation}\label{efold-inflation-2}
    N_\mathrm{f} \simeq - \int_{\phi_\mathrm{c}}^{\phi_\mathrm{f}} \frac{3H^2}{\partial_\phi V} d\phi = \frac{3}{V_0 n \left(2-\frac{1-\beta}{2-\beta}n\right)}  A^{\frac{1}{2-\beta}} \left(\phi_\mathrm{c}^{2-\frac{1-\beta}{2-\beta}n} - \phi_\mathrm{f}^{2-\frac{1-\beta}{2-\beta}n}\right) \ .
\end{equation}
Here, $\phi_\mathrm{c}$ is the scalar field at the beginning of inflation (i.e., at $N=0$) which is considered to be the horizon-crossing instant of the large CMB scale mode ($\sim$0.05~$\mathrm{Mpc}$$^{-1}$). By using Equations~(\ref{ConditionEpsInf}) and~(\ref{efold-inflation-2}), one can easily obtain the expression for $\phi_\mathrm{c}$ as follows:
\begin{equation}\label{phi-c}
    \phi_\mathrm{c} = \left[A^{-\frac{1}{2-\beta}}\left(\frac{nV_0}{3}\left(2-\frac{1-\beta}{2-\beta}n\right)N_f+\frac{V_0n^2}{12(2-\beta)}\right)\right]^\frac{1}{2-\frac{1-\beta}{2-\beta}n} \ .
\end{equation}
Clearly $\phi_\mathrm{c} > \phi_\mathrm{f}$, due to the constraint in Equation~(\ref{constraint}), as expected. Substituting the above form of $\phi_\mathrm{c}$ into Equation~(\ref{InfParPhi}) along with a little bit of simplification lead to the slow roll parameters, evaluated at the instant of horizon crossing, as
\begin{align}\label{S3nrs}
    \epsilon(\phi_c) &= \frac{n}{n + 4(2-\beta)\left(2-\frac{1-\beta}{2-\beta}n\right)N_\mathrm{f}} \\
    \eta(\phi_c) &= \frac{(7-5\beta)n-4(2-\beta)}{n + 4(2-\beta)\left(2-\frac{1-\beta}{2-\beta}n\right)N_\mathrm{f}} \ ,
\end{align}
respectively. Consequently, the scalar spectral index and the tensor-to-scalar ratio are obtained as
\begin{eqnarray}
    n_s&=&1-6\epsilon+2\eta\bigg|_{\phi=\phi_c} = 1 + \frac{2(4-5\beta)n-8(2-\beta)}{n + 4(2-\beta)\left(2-\frac{1-\beta}{2-\beta}n\right)N_\mathrm{f}} \ ,\nonumber\\
    r&=&16\epsilon\bigg|_{\phi=\phi_c} =  \frac{16 n}{n + 4(2-\beta)\left(2-\frac{1-\beta}{2-\beta}n\right)N_\mathrm{f}}\ .
    \label{ns-r-gen}
\end{eqnarray}
Here, we compute the observable indices at the instant of horizon crossing of the large-scale CMB modes on which we are interested to corroborate the theoretical predictions with the Planck 2018 data. According to the Planck result, the $n_s$ and $r$ are constrained by Equation~(\ref{constraint-ns-r}).
From Equation~(\ref{ns-r-gen}), it may be noted that $n_s$ and $r$ is influenced by the entropic parameter $\beta$ and that the dependence by the potential is given by the exponent $n$. {Moreover, $n_s$ and $r$ also depend on the inflationary e-folding number $N_\mathrm{f}$. Actually, the other entropic parameters, in particular $\alpha_\mathrm{\pm}$ and $\gamma$, are packed within $A$ (see Equation~(\ref{A})) and these arise in the final expression of the observable indices through $N_\mathrm{f}$. One can easily choose $\alpha_\mathrm{\pm}$ and $\gamma$ in such a way that $N_\mathrm{f}$ becomes around $\approx$60. Therefore the important constraint that we need to take care is on the parameter $\beta$.} In Figure \ref{fig:3ParRegionn1} and Figure \ref{fig:3ParRegionn2}, we plot the region of validity of $\left\{n_s,r\right\}$ (in respect to the Planck 2018 constraint) in the $\beta\textrm{-}N_\mathrm{f}$ space for the case $n=1$, and the case $n=2$ respectively. 
The presence of an intersection between $\beta$ and $N_\mathrm{f}$ (around $N_\mathrm{f} = 60$) shows the possibility of the agreement of the four-parameter generalized entropic inflation with the Planck data. In both the figures, the gray shadowed region, corresponding to $\beta > 2$, is not acceptable since it leads to negative values of $\epsilon$ (as demonstrated after Equation~(\ref{InfParPhi})).
 
 Thus as a whole, the power law inflaton potential (for $n=1$ as well as for $n=2$) turns out to produce a viable inflation with a graceful exit and is also consistent with the Planck data in the case where the apparent horizon has the four-parameter generalized entropy, unlike to the case of the Bekenstein--Hawking entropy, which fails to show the simultaneous compatibility of the inflationary indices with the observation.

\vspace{-6pt}

\section{Conclusions}\label{sec-conclusion}

We examine the status of the simplest quadratic inflaton potential in the realm of entropic cosmology where the entropy of the apparent horizon is given by the four-parameter generalized entropy and the matter fields inside of the horizon is dominated by a minimally coupled scalar field with a $\phi^{n}$ type of potential (where $\phi$ is the scalar field under consideration). Actually the quadratic potential (i.e., for $n=2$) in standard scalar field cosmology fails to produce a viable inflation, in particular, the $\phi^2$ potential is not compatible with the inflationary observables based on the recent Planck 2018 data. Here, it is good to mention that in the language of entropic cosmology, the standard scalar field cosmology is analogous to the case where the horizon has Bekenstein--Hawking entropy, which results in the usual Friedmann equations. However the story becomes different when the apparent horizon acquires the four-parameter generalized type of entropy, in which case, the entropic energy density plays a significant role during the evolutionary course of the universe. The appearance of the generalized entropy actually generates an effective energy density which modifies the Friedmann equations, and consequently the cosmic evolution of the universe, compared to the standard cosmological scenario. As a result, it turns out that with the four-parameter generalized entropy of the apparent horizon, the $\phi^2$ potential results \linebreak  in---(1) an inflation era described by a quasi de-Sitter evolution of the Hubble parameter, which has an exit at around 55-60 e-folding number, (2) the inflationary observable quantities like the spectral index for primordial scalar perturbation and the tensor-to-scalar ratio are simultaneously compatible with the recent Planck data for suitable values of the entropic parameters, (3) for the same parameter values, the the typical energy scale of the universe at the onset of inflation becomes of the order $\sim 10^{-4}$ (in Planck units). Therefore, this work clearly shows that the $\phi^2$ potential can be made back into the inflationary scenario provided the apparent horizon has the four-parameter generalized type of entropy.

{Regarding the four-parameter generalized entropy, a valid question may be raised about its uniqueness, in particular, whether the $S_\mathrm{4}$ in Equation~(\ref{intro-2}) is the unique four-parameter generalized entropy or one may construct another four-parameter entropy function that also accommodates all the known entropies. However the uniqueness property of $S_\mathrm{4}$ is out of the motivation of this work, and thus we expect to study it in some future work.}
\vspace{6pt}

\authorcontributions{Conceptualization, Tanmoy Paul; Methodology, Tanmoy Paul; Formal analysis, Simone D'Onofrio; Investigation, Tanmoy Paul; Writing---original draft, Simone D'Onofrio; Writing---review and editing, Sergei Odintsov and Tanmoy Paul; Visualization, Sergei Odintsov and Tanmoy Paul; Supervision, Sergei Odintsov. 
}

\funding{This work was partially supported by MICINN (Spain), project PID2019-104397GB-I00 and by the program Unidad de Excelencia Maria de Maeztu CEX2020-001058-M, Spain (S.D.O). This work is funded by MCIN/AEI/10.13039/501100011033 and FSE+, reference PRE2021-098098 (S.O). 
}

\dataavailability{Data are contained within the article. 
} 

\acknowledgments{~ 
}

\conflictsofinterest{The authors declare no conflict of interest. 
}


\appendixtitles{yes} 
\appendixstart
\appendix
\section[\appendixname~\thesection]{Generalized Entropic Inflation with Varying Entropic Parameters}\label{sec-appendix}

In this section, we will show that, beside the consideration of matter fields inside of the horizon, one can get a proper inflation (with a graceful exit) in the context of entropic cosmology by considering the entropic parameters to vary with the cosmic expansion of the universe. In particular, we will concentrate on the three-parameter and the four-parameter generalized entropy, namely the $S_\mathrm{3}$ and the $S_\mathrm{4}$, where the parameter $\gamma$ is considered to vary with the cosmic time. Thus $\gamma = \gamma(N)$ where $N$ represents the e-fold number. The running behavior of $\gamma$ can be described by quantum gravity as, actually, in the sector of gravity, the degrees of freedom may increase if the spacetime fluctuates at high energy scales. 
In the absence of matter fields inside of the apparent horizon, i.e, $\rho = p = \Lambda = 0$, and for $\gamma = \gamma(N)$, the thermodynamic law on the apparent horizon states:
\begin{equation}
    0=T dS_g(S,\gamma(N)) = \frac{\partial S_g}{\partial S} dS +  \frac{\partial S_g}{\partial \gamma} \gamma'(N) dN \ ,
\end{equation}
where a prime indicates the differentiation with respect to $N$, and $S_\mathrm{g} = \left\{S_\mathrm{3},S_\mathrm{4}\right\}$. Taking into account the dependence of the entropy on $\gamma$ and defining $\sigma(N)\equiv\gamma'/\gamma$ we find
\begin{equation}
    \frac{1}{S_\mathrm{g}}\frac{\partial S_\mathrm{g}}{\partial S} dS = \frac{\partial \ln{S_\mathrm{g}}}{\partial S} dS =\sigma(N) dN \ ,
\end{equation}
integrating which we find
\begin{equation} \label{ViableInfFried}
    S_\mathrm{g} = \exp{\left(\int_0^N\sigma(N')dN'\right)} \ .
\end{equation}
At this stage, we will take a certain form of $\gamma(N)$ as follows:

\begin{adjustwidth}{-\extralength}{0cm}
\begin{equation}\label{sigma}
    \gamma(N) = \left\{\begin{aligned}
        &\gamma_\mathrm{0} \textstyle\exp{\left(\int_N^{N_f}\sigma(N')dN'\right)} \hspace{0.7cm} &&N \leq N_f\\
        &\gamma_\mathrm{0} &&N > N_f
    \end{aligned}
    \right. ,\hspace{1.cm} \text{with} \hspace{1.cm}  \sigma(N) \equiv \sigma_\mathrm{0} + e^{-\left(N_\mathrm{f} - N\right)} \ ,
\end{equation}
\end{adjustwidth}
where $\sigma_\mathrm{0}$ is a constant. The second term in the expression of $\sigma(N)$ becomes effective only when $N \approx N_\mathrm{f}$ , i.e., near
the end of inflation. The term $e^{-\left(N_\mathrm{f} - N\right)}$ in the $\sigma(N)$ is actually considered to ensure an exit from inflation era and
thus proves to be an useful one to make the inflationary scenario viable. In the case of the three-parameter entropy,  Equation~(\ref{ViableInfFried}) leads to the relation
\begin{equation}
    \left( 1+\frac{\alpha\pi}{\beta G H^2}\right)^\beta-1 = \exp{\left(\int_0^N\sigma(N')dN'\right)} \ ,
\end{equation}
which for $GH^2\ll 1$ gives the asymptotic solution for the Hubble parameter
\begin{equation}\label{HParInf3}
    H(N)= 4\pi M_{Pl}\sqrt{\frac{\alpha}{\beta}}\left(1+\exp{\left(\int_0^N\sigma(N')dN'\right)} \right)^{-\frac{1}{2\beta}} \ .
\end{equation}
In order to apply the standard inflationary analysis we compute the slow-roll parameter in terms of the e-fold parameter $\epsilon(N)$ as
\begin{equation}
    \epsilon(N) \equiv - \frac{H'(N)}{H(N)} = \frac{1}{2\beta} \frac{\sigma(N)}{\left[1+\exp{
    \left(-\int_0^N\sigma(N')dN'\right)}\right]} \ ,
\end{equation}
and
\begin{equation}
    \frac{\epsilon'(N)}{\epsilon(N)} = \frac{e^{-(N_f-N)}}{\sigma(N)} + \frac{\sigma(N)}{1+\exp{\left(\int_0^N\sigma(N')dN'\right)}} \ ,
\end{equation}
where we used the explicit expression of Equation~(\ref{sigma}) to simplify the result.
In order to impose some constraints on the model parameters we compute various observable indices
such as the primordial curvature perturbation $n_s$ and the tensor-to-scalar ratio $r$ obtained as
\begin{equation}
    n_s = 1-2\epsilon-2\frac{\epsilon'}{\epsilon}\bigg|_{N=0} \quad \text{and} \quad r = 16\epsilon\bigg|_{N=0} \ .
\end{equation}
We compute these indices in the instant of horizon crossing of the large scale CMB modes, which corresponds to the beginning of inflation $N=0$.

\begin{figure}[H]
    \includegraphics[scale=0.8]{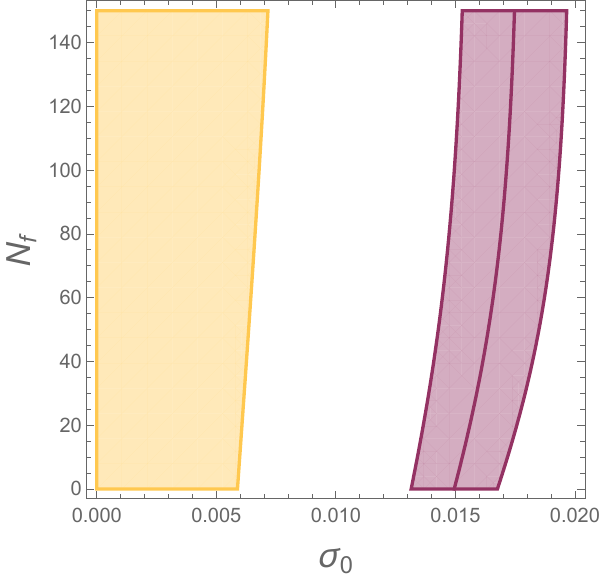}
    \caption{Region of validity for $n_s$ (blue) and $r$ (orange) in case of the three-parameter generalized entropy with varying parameters (by using Equation~(\ref{ns-r-3entropy})) in respect to the Planck data.}
    \label{fig:3ParRegion}
\end{figure}

In the case of the three-parameter entropy with Hubble parameter of Equation~(\ref{HParInf3}) along with $\sigma(N)$ defined in Equation~(\ref{sigma}) the observable indices have the expressions
\begin{eqnarray}
    n_s = 1 -\left( 1+\frac{1}{2\beta}\right)\left(\sigma_0+e^{-N_f}\right) - \frac{2}{1+e^{N_f}\sigma_0}~~~~~~~~\textrm{and}~~~~~~~
    r = \frac{4(\sigma_0+e^{-N_f})}{\beta} \ .
\end{eqnarray}
Since the e-fold number of inflation $N_\mathrm{f}$ will be taken to be $\sim$55--60 we will neglect the terms $e^{-N_\mathrm{f}}$. We can then impose an end to the inflation era with the condition
\begin{equation}
    \epsilon(N_\mathrm{f}) = \frac{1+\sigma_\mathrm{0}}{2\beta(1+e^{-1-N_\mathrm{f}\sigma_\mathrm{0}})} = 1 \ ,
\end{equation}
from which we obtain an inter-relation between $\beta$ and $\sigma_\mathrm{0}$ that can then reduce the number of parameters inside the observable indices. This substitution gives
\begin{eqnarray}\label{ns-r-3entropy}
    n_s = 1-\frac{\sigma_\mathrm{0}\left(2+\sigma_\mathrm{0}+e^{-1-N_\mathrm{f}\sigma_\mathrm{0}}\right)}{1+\sigma_\mathrm{0}} \label{S3n}~~~~~~~~~~\textrm{and}~~~~~~~~~r = \frac{8\sigma_\mathrm{0}\left(1+e^{-1-N_\mathrm{f}\sigma_\mathrm{0}}\right)}{1+\sigma_\mathrm{0}} \label{S3r} \ ,
\end{eqnarray}
which shows that both the $n_s$ and $r$ depend on $\sigma_\mathrm{0}$ and $N_\mathrm{f}$. Since these observable indices are functions of only two variables we can directly check their validity in respect to the Planck 2018 data. In Figure~\ref{fig:3ParRegion} we plot the region of validity of $\left\{n_s,r\right\}$ given in Equation~(\ref{ns-r-3entropy}) in 
 $\sigma_0\textrm{-}N_\mathrm{f}$ space. As we can see from Figure~\ref{fig:3ParRegion}, there is no overlapped region between $\sigma_\mathrm{0}$, and thus the three-parameter generalized entropy with varying parameters does not provide a inflationary scenario that is compatible with the Planck data.

\vspace{-6pt}

\begin{figure}[H]
    \includegraphics[scale=0.8]{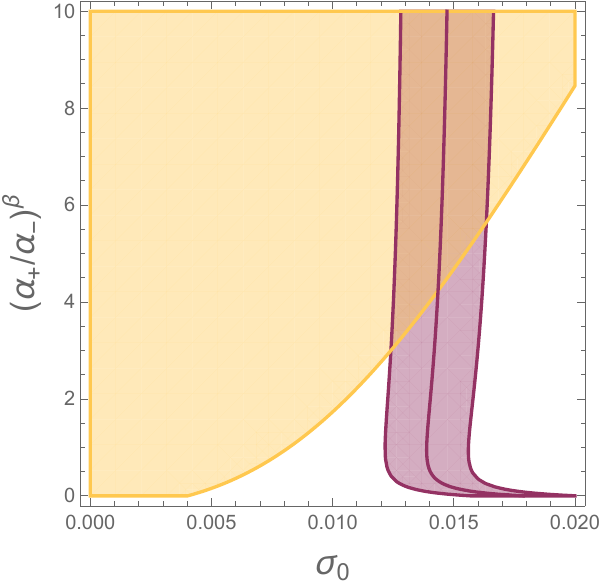}
    \caption{Region of validity for $n_s$ (blue) and $r$ (orange) in case of the four-parameter generalized entropy with varying parameters (by using Equations~(\ref{ns-4entropy}) and~(\ref{r-4entropy})) in respect to the Planck data. Here, we take $N_\mathrm{f} = 58$.}
    \label{fig:4ParRegion}
\end{figure}

In the case of the four-parameter entropy, Equation~(\ref{ViableInfFried}) leads to
\begin{equation}
    \left( 1+\frac{\alpha_+\pi}{\beta G H^2}\right)^\beta-\left( 1+\frac{\alpha_-\pi}{\beta G H^2}\right)^{-\beta} = \exp{\left(\int_0^N\sigma(N')dN'\right)} \ ,
\end{equation}
in the limit $GH^2 \ll 1$ we can safely neglect the plus one in parenthesis and solve the equation for $H(N)$

\begin{adjustwidth}{-\extralength}{0cm}
\begin{equation}
    H(N) = 4\pi M_{Pl}\sqrt{\frac{\alpha_+}{\beta}}\left[\frac{1}{2}\textstyle\exp{\left(\int_0^N\sigma(N')dN'\right)}\left(1+\sqrt{1+4\left(\frac{\alpha_+}{\alpha_-}\right)^\beta \exp{\left(-2\int_0^N\sigma(N')dN'\right)}}\right)\right]^{-\frac{1}{2\beta}} \ .
\end{equation}
\end{adjustwidth}
For this Hubble parameter the slow roll parameter is
\begin{equation}
    \epsilon(N) = \frac{1}{2\beta} \frac{\sigma(N)}{{\sqrt{1+4\left(\frac{\alpha_+}{\alpha_-}\right)^\beta\exp{\left(-2\int_0^N\sigma(N')dN'\right)}}}} \ ,
\end{equation}
and
\begin{equation}
    \frac{\epsilon'(N)}{\epsilon(N)} = \frac{e^{-(N_f-N)}}{\sigma(N)} + \frac{\sigma(N)}{1+\frac{1}{4}\left(\frac{\alpha_+}{\alpha_-}\right)^{-\beta}\exp{\left(2\int_0^N\sigma(N')dN'\right)}} \ .
\end{equation}
Again we will impose the condition on the end of inflation as
\begin{equation}
    \epsilon(N_f) = \frac{1+\sigma_0}{2\beta \sqrt{1+4\left(\frac{\alpha_+}{\alpha_-}\right)^{\beta}e^{-2-2N_f\sigma_0}}}=1 \ ,
\end{equation}
which leads to an interrelation between the parameters $\left(\alpha_+/\alpha_-\right)^\beta$ and $\sigma_\mathrm{0}$. Using such a relation, we obtain the observable indices in the context of four-parameter generalized entropy with varying parameters as follows:
\begin{eqnarray}
    n_s = 1 - \frac{2\sigma_\mathrm{0}\sqrt{1+4\left(\frac{\alpha_+}{\alpha_-}\right)^{\beta}e^{-2-2N_\mathrm{f}\sigma_\mathrm{0}}}}{\left(1+\sigma_\mathrm{0}\right)\sqrt{1+4\left(\frac{\alpha_+}{\alpha_-}\right)^{\beta}}}-\frac{8\sigma_\mathrm{0}\left(\frac{\alpha_+}{\alpha_-}\right)^{\beta}}{1+4\left(\frac{\alpha_+}{\alpha_-}\right)^{\beta}} \ ,
    \label{ns-4entropy}
    \end{eqnarray}
    and
\begin{eqnarray}
    r =  \frac{16\sigma_\mathrm{0}\sqrt{1+4\left(\frac{\alpha_+}{\alpha_-}\right)^{\beta}e^{-2-2N_\mathrm{f}\sigma_\mathrm{0}}}}{\left(1+\sigma_\mathrm{0}\right)\sqrt{1+4\left(\frac{\alpha_+}{\alpha_-}\right)^{\beta}}} \ ,
    \label{r-4entropy}
\end{eqnarray}
respectively. Equations~(\ref{ns-4entropy}) and~(\ref{r-4entropy}) clearly indicate that both the $n_s$ and $r$ depend on $\sigma$, $\left(\alpha_+/\alpha_-\right)^\beta$ and $N_\mathrm{f}$.
For a fixed $N_\mathrm{f}$, particularly around $N_\mathrm{f} = 55\textrm{--}60$, which is consistent with the resolution of the horizon problem, we find the constraints imposed for these variables in order to obtain viable inflation \cite{Planck:2018jri}. The validity of these constraints is shown by the intersection of the two regions in Figure~\ref{fig:4ParRegion} for $N_\mathrm{f} = 58$.

Thus, as a whole, it turns out that in the context of entropic cosmology with varying entropic parameters, the four-parameter generalized entropy can provide a viable inflation consistent with the Planck data, unlike the three-parameter generalized entropy that fails to show the simultaneous compatibility of primordial inflationary indices with the Planck result.
\clearpage

\begin{adjustwidth}{-\extralength}{0cm}

\reftitle{References}

\PublishersNote{}
\end{adjustwidth}

\end{document}